\documentclass[aps,amsmath,amssymb,showpacs,showkeys]{revtex4}

\begin{document}

\title{ Long wavelength properties of electron-TO-phonon interactions \\
        in polar crystals }

\author{Aleksandr Pishtshev}
\email[]{E-mail: ap@eeter.fi.tartu.ee}
\affiliation{Institute of Physics, University of Tartu, Riia 142, 51014 Tartu, Estonia}

\begin{abstract}
\noindent
Theoretical analysis dealing with the interaction of electrons with the polar
long-wavelength transverse optical (TO) vibrations is presented.
The theory is based on the model of a polar crystal with classical potentials,
which takes into account the electronic polarizability effects.
A significant enhancement of the strength of the electron-TO-phonon interaction
in ferroelectrics is found. A microscopic justification of this effect is given.
A bridge that relates the interaction of electrons with the polar long-wavelength TO modes
of the lattice vibrations to the long-range dipole-dipole interaction is established.
As an application of our analysis, a new equation representing the relationship
between the electron-TO-phonon interaction constant and material parameters is obtained.
\end{abstract}
\keywords{Electron-phonon interaction; optical lattice vibrations; polar crystals}
\pacs{63.20.-e; 63.20.kd; 77.84.-s}

\maketitle

%
\section{Introduction}
In the lattice dynamics of a polar crystal, there are two characteristic features,
which give evidence that the interaction of electrons with the polar long-wavelength
optical modes should be necessarily taken into account.
The first feature is related to the existence of dipole moments associated with optical
vibrational modes. The second one concerns with the difference in long-range fields
given by longitudinal and transverse optical modes at the zone center~\cite{BornM}.
In general, the long-wavelength optical vibrations in polar crystals are responsible
for the presence of an electric field~\cite{BornM}.
As it is well-known, the generation of this field results from the displacement
of the ions and is modified by electronic polarizabilities.
This modifies the forces and affects the optical phonon frequencies.
On the other hand, the anomalously large Born effective charges,
which give giant LO-TO splittings in ferroelectric compounds~\cite{Zhong},
are associated with the existence of an anomalously large 
dipole-dipole interaction~\cite{Ghosez}.
The long-wavelength optical phonons, therefore, appear to be a key factor
for relating electronic and lattice (structural) properties.

It is important to emphasize that electron-phonon coupling has the fundamental
significance in the understanding of various physical properties of polar
compounds (e.g.,~\cite{Mechelen,Devreese}).
However, to the best of our knowledge, most comprehensive information
regarding the interactions between electrons and optical phonons is primarily concerned
with the theoretical and experimental studies related to longitudinal optical (LO) phonons
(e.g.,~\cite{Mechelen,Devreese,Harrison,Cardona2,Licari,Alexandrov} and references therein).
As regards transverse optical (TO) vibrations, with only a few
exceptions~\cite{Vinecki,Kristoffel6}, no attempts have been made to consider
consistently the problem of determining the electron-TO-phonon (el-TO-ph) interactions
in polar crystals with due account for their particular polar properties.
This question is of considerable practical
importance for the vibronic theory (e.g.,~\cite{Kristoffel1,Kristoffel2,Bersuker1,Bersuker2}
and references therein) in which the coupling between electrons and the zone-center TO vibrations
is the driving force of a ferroelectric instability, and where the strength of this coupling
should be especially strong~\cite{Pishtshev1}. Moreover, the new developments of
the vibronic theory concerning different properties of oxide ferroelectrics and
multiferroics~\cite{Kristoffel2,Bersuker2,Girshberg1,Konsin3,Konsin4} increased
the interest in the understanding of the nature of~the~el-TO-ph~interactions.
Thus, this raises the issue of very little theoretical knowledge about
the el-TO-ph interaction constants in polar compounds, and the present paper
is aimed at filling this gap.

In this paper, using characteristic parameters of a polar crystal in an explicit way,
we study the main features of the el-TO-ph interaction.
The novel aspects of our work are based on a first-principles treatment of the role
of the el-TO-ph interaction.  
In particular, we demonstrate that, in polar materials, the el-TO-ph coupling
is essentially influenced by the features of the Coulomb interaction between electrons
and the lattice ions.
We also give a microscopic justification of a significant enhancement 
of the el-TO-ph interaction strength in ferroelectrics compounds.
Within a first-principles methodology, we show how to link the interaction of electrons
with the polar long-wavelength TO phonons to the long-range dipole-dipole interaction.
This, in turn, provides a nontrivial relationship between the el-TO-ph interaction constants
and the macroscopic material parameters, such as dipole oscillator strengths
and the forbidden gap. The relationship gives us
a practical framework for estimations of the values of the el-TO-ph interaction constants
in a polar material for which no data have been available so far.

In the more general context of understanding displacive structural instabilities
our results can be used to probe deeper into the nature of the ferroelectric phase transition,
and, in particular, to discover what factors related to material properties
may be responsible for the strength of the el-TO-ph interaction in polar insulators.
%
\section{Theoretical modeling} 
\subsection{Basic set-up of physical model and the main features}
As our focus is the el-TO-ph interaction, we shall be interested in
finding the change in the electronic charge density induced by a polar TO vibration
of a long wavelength.
In order to consider the problem at the atomic level, we represent a polar crystal by
a model based on classical potentials and contributions of the electronic polarization
effects through the relative displacements of the electronic shells~\cite{Vinecki}
(see also~\cite{Sepliarsky1,Wilson}).
Within the standard procedure of the dipole approximation
(expanding to first order in the ionic and electronic displacements,
entering into the~${\bf q}$-representation, and going to the normal coordinates
$u_{{\bf q}j}$), the perturbation of the host crystal due to interactions
with the polar lattice TO vibrations (${\bf q}j$) is given by:

\begin{equation}\label{ei potential1}
 \delta U_{ext}^{tr}({\bf r}) =
N^{-1/2} \sum_{ {\bf q},j  } V_{{\bf q}j}({\bf r})\,
u_{{\bf q}j} 
\end{equation}
where
$V_{{\bf q}j}({\bf r})= v \, {\bf P}({\bf q}j)\cdot{\bf F}_{{\bf q}}({\bf r})\,$,
\begin{equation}\label{ei potential1B}
{\bf F}_{{\bf q}}({\bf r}) =
\,-\, ie \, \frac{4\pi}{v} \, {\sum_{{\bf G}\neq0}} \,
{\frac{{\bf G}}{{\vert{\bf q}+{\bf G}\vert}^2}} \,
{ e^{i\,({\bf q}+{\bf G})\,{\bf r}} }  \, ,
\end{equation}
${\bf P}({\bf q}j)$ is an amplitude of the dipole polarization
(dipole moment per unit volume) associated with the polar TO mode~\cite{BornM},
${\bf G}$ denotes the reciprocal lattice vector,
$v$ is the volume of the unit cell, and $N$ is their number.
The derivation of Eq.~(\ref{ei potential1}) accounts also for the fact
that in the long-wavelength limit one can take the corresponding Fourier transformation
of the Coulomb potential to be independent of ion's position in the unit cell.
The accuracy of this approximation was investigated previously in~\cite{Vinecki,Kristoffel6}.

Equation~(\ref{ei potential1}) is exactly the one that we shall use
to study the interaction of electrons with the polar TO lattice vibrations
in the long-wavelength limit. Before presenting the results themselves, 
let us emphasize the most important key points which 
characterize the model as a consistent, systematic and workable framework.

The first is that Eq.~(\ref{ei potential1}) accounts for the changes in ion properties
that are caused by changes
in its environment, and, hence, in accordance with~\cite{Wilson} includes not only pairwise
electron-ion interactions but also many-body effects.
This follows from the accounting for an electronic contribution (${\bf P}^{(e)}({\bf q}j)$)
in the dipole polarization:
${\bf P}({\bf q}j) = {\bf P}^{(i)}({\bf q}j) + {\bf P}^{(e)}({\bf q}j)$.
This contribution is connected with the the distortion of the electronic
charge distribution (e.g.,~\cite{Hardy}).
The corresponding ionic contribution (${\bf P}^{(i)}({\bf q}j)$) is related with
the displacements of the ions participating in the TO vibrations~\cite{BornM}:
${\bf P}^{(i)}({\bf q}j)= v^{-1}
{\sum}_{s} ({ Z_s }e/{ \sqrt{{M_s}} } \,) {\bf w}(s,{\bf q}j)$
where ${Z_s}e$ denotes the effective charge~\cite{Cochran} of an ion with the mass ${M_s}$
placed in the cite $s$, and ${\bf w}(s,{\bf q}j)$ is the polarization vector of the normal
TO vibration. 
At ${\bf q}=0$, the both polarization components satisfy the following
relations~\cite{BornM,Vinecki}:
\begin{eqnarray}\label{polvector2}
{\bf P}^{(i)}(0j) =\,
\frac{3}{{\epsilon}_{\infty}+2}\, {\bf P}(0j) \, , \,\,
{\bf P}^{(e)}(0j) =\,
\frac{{\epsilon}_{\infty}-1}{{\epsilon}_{\infty}+2}\, {\bf P}(0j) 
\end{eqnarray}
where ${\epsilon}_{\infty}$ is the electronic (static-high-frequency) dielectric permittivity. 

Attention should next be directed to the long-wavelength nature of the quantity
${\bf F}_{{\bf q}}({\bf r})$ standing in Eq.~(\ref{ei potential1}).
On the one hand, in a polar crystal it defines the electric field at a point ${\bf r}$
inside the bulk as a part of the internal field associated with
the TO phonon displacement~\cite{Vinecki} (we recall that the microscopic
electric field excited in the crystal has components ${\bf q}+{\bf G}$
for all reciprocal-lattice vectors). On the other hand,
at ${\bf q}=0$, 
$\left.{\bf F}_{{\bf q}}({\bf r})\right|_{{\bf q}=0}={\bf F}_{0}({\bf r})$
can be represented as (${\bf l}$ denotes lattice vectors):
\begin{equation}\label{transversefield1}
{\bf F}_{0}({\bf r})\,=\,-\,
{\bf \nabla}\left({ {\sum_{\bf l}}\,
\frac{e}{\left|{\bf r}-{\bf l}\right|} }\right) \,=\,-\,
{\bf \nabla} \phi_{e}({\bf r}) \, .
\end{equation}
It is seen from Eq.~(\ref{transversefield1}) that, in accordance
with~\cite{Kholopov1,Rozzi}, the quantity ${\bf F}_{0}$
is a field generated by the Coulomb electrostatic potential
$\phi_{e}({\bf r})$ which is connected (at given boundary conditions)
through the Poisson equation with a charge distribution
${\delta}{\rho}_e({\bf r})=e{\sum}_{\bf l}{\delta}({\bf r}-{\bf l})$.
This result, in turn, suggests a macroscopic characterization for the quantity
$V_{{\bf q}j}({\bf r})$ given in Eq.~(\ref{ei potential1}) as follows:
it comes in the form of
an electrostatic contribution which interrelates the dipole polarization
associated with the polar long-wavelength TO mode
and
the electric field caused by the reaction of the electron subsystem
to the relevant {\textquotedblleft}external{\textquotedblright} perturbation.
In this connection, we also note that both the resulting shift of the electron density and
the effect of the dipole polarization appear to be clear signatures of the strength of
the el-TO-ph interaction. 

The third important point is that interaction in the form~(\ref{ei potential1})
may be responsible for the softening of the lattice subsystem.
Clearly, this is the case of the vibronic theory~\cite{Kristoffel1,Bersuker1}.
Following closely Ref.~\cite{Hargittai}, we observe that local change in the electron
density distribution will generate an imbalance with respect to the corresponding density
of the positive background. As a result, there will be a restoring force
that will tend to compensate this density imbalance. 
As it argued in~\cite{Hargittai}, such asymmetry effect will initiate  distortions
of the ions and the redistribution of charge, eventually leading to a decrease
in the total energy, and to a match between the symmetry of the ionic arrangement and 
that of the electron density distribution.
%
\subsection{Model for electron-TO phonon interaction}
%
The starting point is a generic model of an insulator
consisting of the occupied valence and empty conduction bands
separated by a forbidden gap.
Keeping in mind correctness in the description of an insulated state,
we assume that the wave functions ($\psi_{{\sigma}\bf k}$)
and energies ($E_{\sigma}({\bf k})$) are the set of one-particle Bloch eigenfunctions
and eigenvalues corresponding to the band structure, 
which are determined within the framework of DFT-LDA methods
with the correct accounting for the quasiparticle effects
by means of many-body perturbation theory~\cite{Aryasetiawan1}.

Equation~(\ref{ei potential1}), providing a basis for studying
the el-TO-ph interaction, corresponds in the second-quantization
formalism to the Hamiltonian
\begin{equation}\label{hamiltonian_el-ph}
H_{el-ph} =
{N}^{-1/2} \,
{\sum_{\sigma,\sigma^{'}\,j}}\,{\sum_{\bf k,\bf q}}\,
g_{{\sigma}{\sigma}^{'}}({\bf k},{\bf q}j) \,
a^+_{\sigma{\bf k}}a_{\sigma^{'}{\bf k-q}} \, u_{{\bf q}j} \,
\end{equation}
which characterizes the dynamic mixing of an electron state
$\vert\,{\sigma}\,{\bf k}\,${\textgreater}
with another state $|\,{{\sigma}^{'}\bf (\bf{k}-\bf{q})}\,${\textgreater}
caused by the TO phonon.
Here $a^+$ ($a$) are the creation (annihilation) operators
for electronic states 
in the valence and conduction bands ($\sigma, \sigma^{'}$),
$g_{{\sigma}{\sigma}^{'}}({\bf k},{\bf q}j)$~denotes
the matrix elements of the el-TO-ph interaction defined
at the equilibrium high-symmetry configuration of the lattice by
\begin{equation}\label{integral_el-ph0}
g_{{\sigma}{\sigma}^{'}}({\bf k},{\bf q}j) \,=\,
<{\sigma}{\bf k}|\,
V_{{\bf q}j}({\bf r})
|{\sigma}^{'}({\bf k}-{\bf q})>\, .
\end{equation}
Expression~(\ref{integral_el-ph0}) involves both
interband (${{\sigma}\neq{\sigma}^{'}}$) and intraband (${{\sigma}={\sigma}^{'}}$)
matrix elements. 
In the long-wavelength limit, the intraband matrix elements vanish due to inversion symmetry
and, correspondingly, the coupling of electrons and TO lattice vibrations
is completely characterized by the interband matrix elements.

In the context of first-principles features/attributes belonging to the model,
it is worth to emphasize the following aspects:
(i) The electronic states are periodic in space and delocalized over
the entire unit cell; in systems with a gap
infinitesimal displacements affect all the states uniformly~\cite{Stechel},
(ii) the valence and conduction (excited) states can mix under these 
displacements~\cite{Bersuker2},
(iii) the energetics of the interaction of electrons with the long-wavelength
polar TO lattice vibrations
is strongly affected by energetics of the lattice electronic distribution,
(iv) the effect of the dipole polarization is to change the 
center of force between a "shell" of an ion's electrons and 
the other force centers in the crystal~\cite{Hardy2}, 
(v) like the electron-gas model of Gordon and Kim~\cite{GKmodel}, the corresponding
interaction energy is a function of changes in the electron charge densities, and
(vi) as seen from Eq.~(\ref{transversefield1}), these changes are represented
by distortions of charge density owing to electric field gradients at certain lattice sites
of a polar crystal.

Our intermediate goal here has been to offer a detailed, microscopically realistic
framework that could be used for quantitative investigations of the el-TO-ph
interaction in a polar crystal. In order to achieve the objectives of the present work 
further study in this direction is divided into two main steps:
The first is a comparative first-principles analysis of the contribution
of Eq.~(\ref{hamiltonian_el-ph}) into the lattice dynamics of TO lattice vibrations;
the second handles the interband matrix elements, and develops formulas
allowing us to estimate el-TO-ph interaction strengths in terms of macroscopic material
parameters for a wide range of polar dielectrics.
%
\section{Electron-phonon interaction and TO lattice vibrations} 
%
\subsection{On the dynamics of TO lattice vibrations}
We begin the present section by reviewing the main properties of 
the electronic contribution into the dynamics of lattice vibrations.
Then we calculate the relevant electronic contribution given by Eq.~(\ref{hamiltonian_el-ph}).
Our task is to compare this contribution with the corresponding results of microscopic
lattice dynamics~\cite{Sham2,Kvyatkovskii2,Kvyatkovskii3} with regard to the long-wavelength
polar TO phonons. For this purpose, within the adiabatic approximation, we consider the dynamics
of the TO vibrations in terms of the matrix elements of the el-TO-ph interaction
determined by Eq.~(\ref{integral_el-ph0}).

A key feature of the electron subsystem is that it provides a link between an electronic energy
and an equilibrium geometry of a crystalline structure (this link in turn depends upon the
electronic contributions to the total energy).
The electronic contributions to the dynamical matrix of a nonmagnetic crystal arise from changes
in the electronic charge density due to the presence of the electron-ion potential in 
a system (e.g.,~\cite{Kristoffel1,Bersuker1,Sham2,Baroni2,Baroni1}).
This implies that, by studying the lattice dynamics, one can acquire fundamental information on
the role of the electron-phonon interactions in a given material.

Since the real part of the self-energy of the TO vibrational mode
specifies the deviation of the unperturbed phonon states upon adiabatic variation
of the potential connected with external perturbation,
we represent the square of the TO phonon frequency $\Omega_{{\bf q}j}^{2}$
as a sum of the unperturbed term $\widetilde{\omega}^{2}_{{\bf q}j}$ and
the additional term $\Delta\omega^{2}_{{\bf q}j}$ which accounts for el-TO-ph coupling
governed by Hamiltonian~(\ref{hamiltonian_el-ph}):
$$
\Omega^{2}_{{\bf q}j} = \widetilde{\omega}^{2}_{{\bf q}j} + \Delta\omega^{2}_{{\bf q}j} \,.
$$
We expect the {\textquotedblleft}normal{\textquotedblright}
part $\widetilde{\omega}^{2}_{{\bf q}j}$,
which may be regarded as the bare TO phonon frequency
with respect to the el-TO-ph coupling, to be free of softening anomalies.
As usual (e.g.,~\cite{Kristoffel1,Cowley}), it is assumed that
$\widetilde{\omega}^{2}_{{\bf q}j}$ also involves contributions from the phonon-phonon
interactions, which are needed to describe the dependence of the TO phonon frequency
on temperature. 

We can obtain the corresponding electronic contribution $\Delta\omega^{2}_{{\bf q}j}$ in
the following way. The first step is to employ
the linear density response function matrix (the dielectric susceptibility) $\hat{\chi}$
that relates a variation of the electronic density (the charge response) $\delta\hat{{\rho}}$
to the perturbation potential $\delta\hat{v}_{ext}$ as follows~\cite{Dolgov,Louie2}:
\begin{eqnarray}\label{linear-resp1}
\delta\hat{{\rho}}=\hat{\chi}\delta\hat{v}_{ext} \, , \quad
{\hat{\epsilon}}^{-1}=\hat{I}+\hat{v}_{c}\hat{\chi} 
\end{eqnarray}
where ${\hat{\epsilon}}^{-1}$ denotes the inverse dielectric matrix,
$\hat{I}$ is the identity matrix, and $\hat{v}_{c}$ is the bare Coulomb interaction.
Note that perturbation~(\ref{hamiltonian_el-ph}) modifies the ground-state values
and therefore induces $\delta\hat{{\rho}}$.
As stated by the DFT formalism,
both $\hat{\chi}$ and the potential $\delta\hat{v}_{ext}$ itself are unique functionals
of the electronic density.

Since the Coulomb and electron-phonon interactions cause the phonon propagator
to be fully renormalized (e.g.,~\cite{Allen}),
the second step is to represent the real part of the TO phonon self-energy
to the lowest-order in $g^2$ as the corresponding harmonic term,
and to account for the contributions of the Coulomb interaction
among electrons. Using microscopic expressions for matrix elements
$g_{{\sigma}{\sigma}^{'}}({\bf k},{\bf q}j)$ given by Eq.~(\ref{integral_el-ph0}),
by the last step, we write the equation for $\Delta\omega^{2}_{{\bf q}j}$
in the following form:
\begin{equation}\label{softmode-eq2}
\Delta\omega^{2}_{{\bf q}j} =
4\pi v \, {\sum_{\alpha,\beta}}\,B_{\alpha \beta}({\bf q})
\,P_{{\alpha}}({\bf q}j)\,P_{{\beta}}({\bf q}j) 
\end{equation}
where
\begin{equation} \label{softmode-eq2-B}
B_{\alpha \beta}({\bf q}) =  \frac{ v } {4\pi e^{2}} 
{\sum_{{\bf G,G^{'}}{\neq0}}} 
G_{\alpha} \, v_{c}({\bf q+G})\,
\chi_{el}({\bf q+G}, {\bf q+G^{'}})\,
v_{c}({\bf q+G^{'}}) \, G^{'}_{\beta} \, ,  
\end{equation}
\begin{eqnarray}\label{hi-eq1}
{\hat{\chi}}_{el} =
\frac{\hat{\Pi}_{0}}{\hat{I}-\hat{v}_{c}\hat{\Pi}_{0}} \, , \,\,
v_{c}({\bf q+G}) = \frac{4\pi e^{2}}{v\left|{\bf q+G}\right|^{2}} \, ,
\end{eqnarray}
\begin{multline}\label{pol-operator-eq1}
\Pi_{0}({\bf q}+{\bf G}, {\bf q}+{\bf G^{'}}) = 
\,-\, \frac{1}{N} \, {\sum_{\sigma,\sigma^{'}}} {\sum_{\bf k}}\,
\frac{f_{\sigma}({\bf k})-f_{\sigma^{'}}({\bf k+q})}
{E_{\sigma^{'}}({\bf k+q})-E_{\sigma}({\bf k})}  \\ 
{\times} \,
\int \psi^*_{\sigma{\bf k}}\,
e^{i({\bf q}+{\bf G})\,{\bf r}}
\psi_{\sigma^{'}{\bf k-q}}\,d\tau 
\int \psi^*_{\sigma^{'}{\bf k-q}}\,
e^{-i({\bf q}+{\bf G^{'}})\,{\bf r^{'}}}
\psi_{\sigma{\bf k}}\,d\tau^{'}\, .
\end{multline}
Here the matrix elements of the microscopical dielectric susceptibility
of the electron subsystem $\chi_{el}({\bf q+G}, {\bf q+G^{'}})$
are expressed in the random phase approximation
in terms of polarization operator $\hat{\Pi}_{0}$~\cite{Dolgov,Adler,Wiser,Mazin2},
and $f$ are the occupation numbers.
%
\subsection{Consistency with first-principles considerations}
In this and next section, we shall show that our results 
given by Eqs.~(\ref{softmode-eq2})-(\ref{pol-operator-eq1})
are consistent with first-principles considerations.
This will allow us to establish a bridge that relates the interaction of electrons
with the polar long-wavelength TO modes of the lattice vibrations
to the long-range dipole-dipole interaction.

Within the framework of our comparative analysis, it seems to be useful
to take into account three important points:\\
(i) We deal with the long-wavelength TO phonons, i.e., in fact,
with the analytical part of the harmonic force constants.
The remaining non-analytic part only affects the LO phonon frequencies,
and therefore determines the LO-TO splitting at the $\Gamma$ point.\\
(ii) The unique feature of the force constants is that they 
can be decomposed into a sum of two independent terms~\cite{Sham2,Baroni2,Baroni1}:
the direct ionic and electronic contributions.
The electronic contribution, which represents indirect interactions via the electron subsystem,
is the focus of our discussion because involves the detailed information
on the el-TO-ph interaction.\\
(iii) The long-wavelength behavior of the force constants can also be represented in terms of
the balance between the short-range repulsive and long-range Coulomb forces associated
with the short-range and long-range dipole-dipole interactions,
respectively~\cite{Ghosez,Kvyatkovskii2,Kvyatkovskii3,Gonze}.
The long-range contribution is represented by the interplay of the Born transverse effective
charge tensors. It is a key quantity in polar compounds because favors the ferroelectric
distortion, while its direct competitor (the short-range contribution) tends to suppress
the ionic displacements and to provide the stability of the high-symmetry configuration.

Note that, in the context of the present work,
the above pair-wise separabilities of the force constants will be particularly instructive
for comparison purposes. For instance,
by matching different pieces of information contained in the dynamical matrix,
we could recognize relevant aspects of the el-TO-ph interaction.

We start our comparison procedure by noting that within the microscopic
theory of lattice vibrations the Fourier transforms of the harmonic force constants
are expressed  through the matrix elements of the susceptibility function
$\chi_{el}({\bf q+G}, {\bf q+G^{'}})$ (e.g.,~\cite{Sham2,Keating,Pick}).
By performing a comparison of Eqs.~(\ref{softmode-eq2}) and~(\ref{softmode-eq2-B})
with those from~\cite{Sham2}, we can ensure that
the expression for $B_{\alpha \beta}({\bf q})$, which is written in terms of
the microscopic electronic dielectric susceptibilities and the Coulomb potentials,
describes the corresponding electronic
contribution to the square of the TO phonon frequency $\Omega^{2}_{{\bf q}j}$
(i.e., the contribution which uniquely involves the $\chi_{el}({\bf q+G}, {\bf q+G^{'}})$
matrix elements with both ${\bf G}$ and ${\bf G^{'}}$ different from zero). 
Due to the microscopic formulation of the el-TO-ph interaction given by~(\ref{integral_el-ph0}),
this result is general and independent on the particular details of 
the interaction of electrons with the polar long-wavelength TO vibrations.
This implies that fundamental property of $B_{\alpha \beta}({\bf q})$ is that it represents,
in accordance with~\cite{Sham2,Allen,Pick},
the electron mediated part of the long-range Coulomb interaction.
%
\subsection{Relevance to the long-range dipole-dipole interaction}
As the second step of the comparison procedure, it is appropriate to compare
equations~(\ref{softmode-eq2}) and~(\ref{softmode-eq2-B})
with the corresponding results of lattice dynamics
theories~\cite{BornM,Cochran,Sham2,Kvyatkovskii2,Kvyatkovskii3}
related to polar insulators.
For this purpose let us inspect the long-wavelength limit of Eqs.~(\ref{softmode-eq2})
and~(\ref{softmode-eq2-B}) in more detail.

Observe that one of the characteristic features of Eq.~(\ref{softmode-eq2})
represents the interplay of the polarization amplitudes
$P_{{\alpha}}(0j)$ and $P_{{\beta}}(0j)$ 
which, as in~\cite{Kvyatkovskii3}, can be characterized by the second rank tensor
$S_{\alpha \beta}(j) =
4\pi v\, P_{{\alpha}}(0j)P_{{\beta}}(0j)$.
The quantities $S_{\alpha \beta}(j)$
are defined as dipole oscillator strengths for the given zone-centre TO vibration
and correspond to the contributions
of the TO vibrational modes in the infrared (IR) part of the Lorentz model of
the dielectric matrix
${\epsilon}_{\alpha \beta}$ (e.g.,~\cite{Vinecki,Sham2,Cowley2}):
\begin{equation}\label{dm-eq1} 
({\epsilon}_{\omega})_{\alpha \beta} \,=\,
({\epsilon}_{\infty})_{\alpha \beta} \,+\,
{\sum_{j}}\, 
\frac{S_{\alpha \beta}(j)}{\, \Omega^{2}_{0j}{\,-\,}\omega^{2} \,} \, .
\end{equation}
It is easy to show, using the definition of $P_{{\alpha}}^{(i)}(0j)$ together
with~(\ref{polvector2}), that the dipole oscillator strengths
$S_{\alpha \beta}(j)$ can be represented in terms of the TO mode effective charge:
\begin{equation}\label{dm-eq1A} 
S_{\alpha \beta}(j) =\, \frac{4{\pi}e^{2}}{v} \, \left(
\sum_{s} Z_s^{*} \, \frac {w_{\alpha}(s,0j)}{ \sqrt{ {M_s} } } \, \right)
 \left(
\sum_{t} Z_t^{*} \, \frac {w_{\beta}(t,0j)}{ \sqrt{ {M_t} } } \, \right) \, .
\end{equation}
Here the result of each such summation in the round brackets
is called the TO mode effective charge~\cite{Zhong}
(which corresponds to the dipolar activity of the zone-center TO phonons
and is a measure of the intensities of IR-active modes),
$Z_s^{*}$ are the Born transverse dynamical effective charges
defined as~\cite{BornM,Kvyatkovskii2,Ghosez2,Xu}
$Z_s^{*}=\left[({{\epsilon}_{\infty}+2})/3\right]{Z_s}\,$.

After substituting for the product 
$P_{{\alpha}}(0j)P_{{\beta}}(0j)$, Eq.~(\ref{softmode-eq2}) becomes
\begin{equation}\label{softmode-eq3}
\Delta\omega^{2}_{0j} \,=\,
{\sum_{\alpha,\beta}}\, B_{\alpha \beta}(0)\,S_{\alpha \beta}(j) \, 
\end{equation}
where the tensor $B_{\alpha \beta}(0)$ is defined from Eq.~(\ref{softmode-eq2-B})
at ${\bf q}=0$ by
\begin{equation}\label{softmode-eq3-B}
B_{\alpha \beta}(0) =  \frac{ v } {4\pi e^{2}}
{\sum_{{\bf G,G^{'}}{\neq0}}}
G_{\alpha} \, v_{c}({\bf G})\,
\chi_{el}({\bf G}, {\bf G^{'}})\,
v_{c}({\bf G^{'}}) \, G^{'}_{\beta}
\end{equation}
and~$\chi_{el}({\bf G}, {\bf G^{'}})=
{\lim}_{{\bf q}\rightarrow0}\,\chi_{el}({\bf q+G}, {\bf q+G^{'}})$.
By using the representation of the susceptibility
function~$\chi_{el}({\bf G}, {\bf G^{'}})$~in
terms of matrix elements of the microscopic polarizability
tensor~$a_{{\gamma}{\eta}}({\bf G}, {\bf G^{'}})$~\cite{Sinha2},
$$
\chi_{el}({\bf G}, {\bf G^{'}})= - \sum_{\gamma \, \eta}\,
G_{\gamma}\,a_{{\gamma}{\eta}}({\bf G}, {\bf G^{'}})\,G^{'}_{\eta} \, ,
$$
Eq.~(\ref{softmode-eq3-B}) becomes
\begin{equation}\label{softmode-eq3-C}
B_{\alpha \beta}(0) = \,-\, \frac {4\pi e^{2}} { v }
{\sum_{{\bf G,G^{'}}{\neq0}}} \, \sum_{\gamma , \eta}\,
\frac {G_{\alpha} \, G_{\gamma}} { \left|{\bf G}\right|^{2} } \, 
a_{{\gamma}{\eta}}({\bf G}, {\bf G^{'}})\,
\frac { G^{'}_{\eta} \, G^{'}_{\beta} } { \left|{\bf G^{'}}\right|^{2} } \, .
\end{equation}

The set of equations~(\ref{softmode-eq3})-(\ref{softmode-eq3-C}),
which we have rigorously derived, allows us to analyse
the relevance of our model to long-range dipole forces.
We shall now prove this relevance by comparing with
the results of the previous theoretical studies of lattice dynamical models
(e.g.,~\cite{BornM,Cochran,Kvyatkovskii3}).

Note first that, in polar insulators, in the limit of zero wave-vector
the displacement of charges from their equilibrium positions creates
dipoles, which interact with long-range forces~\cite{BornM}.
The associated long-range dipole-dipole interaction can be characterized by
the coupling of the Born transverse dynamical effective
charges~$Z_s^{*}\,Z_t^{*}$, i.e., by the product
$P_{{\alpha}}(0j)\,P_{{\beta}}(0j)\,$.
Secondly, this interaction can be characterized by
assigning the dipole polarization an external moment~\cite{Mahan}
${\bf P}(0j)=[({{\epsilon}_{\infty}+2})/{3}]\,{\bf P}^{(i)}(0j)$
with the proportionality coefficient given by the local-field factor
(see Eqs.~(\ref{polvector2})).
Thirdly, following Ref.~\cite{Iishi}, Eq.~(\ref{softmode-eq3}) with the help of
Eqs.~(\ref{dm-eq1A}) and~(\ref{softmode-eq3-C}) can at once be identified
as the long-range Coulomb contribution resulting from the induced dipole interactions
through the electronic polarizability.

Therefore, our comparative analysis shows that
the set of equations~(\ref{softmode-eq3})-(\ref{softmode-eq3-C})
are directly related to the contribution of the long-range Coulomb interaction.

Furthermore, from comparison of our result given by Eq.~(\ref{softmode-eq3})
with the corresponding equations for the TO vibrational modes derived within
the macroscopical formalism~\cite{BornM}, we can find the evaluation of 
the tensor $B_{\alpha \beta}(0)$.
Let us show that $B_{\alpha \beta}(0)$ is expressed as
\begin{equation}\label{softmode-eq3-B1}
B_{\alpha \beta}(0) =
 -\,{\delta}_{\alpha\beta}/{({\epsilon}_{\infty}+2)} \, .
\end{equation}
As a comparison parameter, it is reasonable to choose the product
$P_{{\alpha}}(0j)\,P_{{\beta}}(0j)$
that measures the long-range dipole-dipole interaction~\cite{Mahan}.
With the use of results of~\cite{BornM,Vinecki},
for the case of the cubic (or tetrahedral) lattice symmetry, in dipole approximations,
the counterpart of Eq.~(\ref{softmode-eq3}) can be represented
in terms of the product $P_{{\alpha}}(0j)\,P_{{\beta}}(0j)$ (or $S_{\alpha \beta}(j)$)
as follows:
\begin{equation}\label{softmode-eqDD1}
\Delta\omega^{2}_{0j} =\,-\,
 4\pi v \, {\sum_{\alpha,\beta}}\,
 \frac{ {\delta}_{\alpha\beta} }{ {\epsilon}_{\infty}+2 } \,
P_{{\alpha}}(0j)\,P_{{\beta}}(0j) 
=\,
- \, {\sum_{\alpha,\beta}}\,
\frac{ {\delta}_{\alpha\beta} }{ {\epsilon}_{\infty}+2 } \,
 S_{\alpha \beta}(j) \,.
\end{equation}
Eq.~(\ref{softmode-eq3-B1}) is obtained by a direct comparison of Eqs.~(\ref{softmode-eq3})
and~(\ref{softmode-eqDD1}).

In the last step of our comparison analysis, we find that 
equations~(\ref{softmode-eq3}) and~(\ref{softmode-eq3-B1}) are identical 
to the results derived within the exact accounting of the dipole-dipole interaction
in lattice dynamics~\cite{Kvyatkovskii3}. This identity directly implies that
the product $B_{\alpha \beta}(0) P_{{\alpha}}(0j)P_{{\beta}}(0j)$
represents the regular at the $q = 0$ contribution of the long-range dipole forces.
This allows us to conclude that, within the framework
of TO vibration mode dynamics, the model originally specified by
Eqs.~(\ref{ei potential1}),~(\ref{ei potential1B}),~(\ref{hamiltonian_el-ph})
and~(\ref{integral_el-ph0}) matches the microscopic approach
proposed and developed for polar crystals in~\cite{Kvyatkovskii2,Kvyatkovskii3}.
This is a novel result of principal importance:
we showed and confirmed the correspondence between two descriptions
of dynamics of the zone-centre TO vibrational mode - one in terms
of the el-TO-ph interaction and the other in terms
of the long-range regular at the $q = 0$ part of the dipole-dipole
interaction~\cite{Kvyatkovskii2,Kvyatkovskii3}.
Moreover, Eq.~(\ref{softmode-eq3}) clearly demonstrates the equivalence 
of the interpreting power of both descriptions, and therefore can serve as a link
between the two approaches.

Thus, we have shown above that microscopic formulation of the el-TO-ph interaction
given by~(\ref{integral_el-ph0}) provides results consistent with first-principles
considerations. We finish this section with a few important remarks.
Firstly, it should be noted that, as follows from the foregoing analysis,
the interaction of electrons with the polar zone-centre TO phonons
is directly associated with the long-range Coulomb interaction.
Secondly, the magnitude of the destabilizing contribution ($B_{\alpha \beta}(0)<0$)
determined by Eq.~(\ref{softmode-eq3}) will rise steeply in polar materials.
This will lead to significant softening of the zone-centre TO vibrational mode
frequency $\Omega_{0j}$.
The third remark is concerned with strong sensitivity
of calculated (at ${\bf q}=0$) frequencies~(\ref{softmode-eq3})
to the dipole oscillator strengths:
this is the manifestation of the fact that the zone-centre TO vibrational modes
exceptionally strongly can be coupled to valence band electrons.

Our findings give a clear understanding of the role of the el-TO-ph interaction
in the conversion of $\Omega_{0j}$ to the ferroelectric soft-mode.
Eq.~(\ref{softmode-eq3}) shows how changes in the electron density
(or the changes in the electronic states) induced by the local displacements
are sufficient to yield a considerable softening of the TO phonons around the $\Gamma$ point.
In the context of structural dynamical instability the dipole oscillator strengths
$S_{\alpha \beta}(j)$ as enhancement factors in Eq.~(\ref{softmode-eq3})
can serve as an important macroscopic measure for
probing how close the system may be to the ferroelectric instability.

In the next section, we specify our treatment of the el-TO-ph interaction to the
case of ferroelectric crystals, and derive the corresponding effective Hamiltonian, 
which would characterize the interband scatterings of the bond electrons
due to the zone-center TO phonons.
\section{On electron-TO-phonon coupling in ferroelectrics }
\subsection{Parametrization of electron-TO phonon coupling at $q=0$ }
The microscopic approach considered above is quite general; in the present section, we show
how simplifications allow us to deduce a reduced model of el-TO-ph interaction.

The starting point is Eqs.~(\ref{softmode-eq2})-(\ref{pol-operator-eq1}) in which
we set ${\bf q}=0$.
The main complication in executing the limit ${\bf q}\rightarrow0$ in the matrix elements
of $\hat{\chi}_{el}$ is that one needs to exclude the relevant contributions associated with
the electronic displacements (these contributions are included in the definition of
$g_{{\sigma}{\sigma}^{'}}^{2}$).
This problem is solved in two steps:
(i) by finding a change of the electronic charge density
associated with interaction~(\ref{ei potential1}) and
(ii) by applying Eqs.~(\ref{linear-resp1})
in a similar way as was described in~\cite{Kvyatkovskii2}
regarding the analysis on the internal electric field
effects. Acting in this manner, and taking into account Eqs.~(\ref{polvector2}),
we obtain that in the long-wavelength limit,
the reduced susceptibility $\hat{{\chi}}_{el}^{0}$ turns out to be related
to $\hat{\chi}_{el}$ by the following equation:
\begin{eqnarray}\label{linear-resp2}
{\bf q}\rightarrow0 : \quad
\hat{\chi}_{el} \,=\,
{\frac{3}{{\epsilon}_{\infty}+2}} \, \hat{{\chi}}_{el}^{0} \,.
\end{eqnarray}

With all of these considerations and assuming that in insulators
$(f_{\sigma}({\bf k})-f_{\sigma^{'}}({\bf k}))=1$,
we can write the long-wavelength limit of
Eqs.~(\ref{softmode-eq2}) in terms of the el-TO-ph interaction
$g_{{\sigma}{\sigma}^{'}}({\bf k},0j){\equiv}g^{j}_{{\sigma}{\sigma^{'}}}({\bf k})$
as follows:
\begin{equation}\label{softmode-fr}
\Delta\omega^{2}_{0j} = \,-\,
{\frac{3}{{\epsilon}_{\infty}+2}} \, \frac{1}{N} 
{{\sum_{\sigma{\neq}\,\sigma^{'}}} \, {\sum_{\bf k}}} \,
\frac{{\vert \, g^{j}_{{\sigma}{\sigma^{'}}}({\bf k}) \vert}^2}
{ \,\left|\,{E_{\sigma^{'}}({\bf k})-E_{\sigma}({\bf k})} \right|\, } \, , 
\end{equation}
\begin{multline}\label{softmode-frA}
{\vert \, g^{j}_{{\sigma}{\sigma^{'}}}({\bf k}) \vert}^2 =
\frac{v^{2}}{e^{2}} \,
{\sum_{\alpha,\beta}}\,P_{{\alpha}}(0j)\,P_{{\beta}}(0j)
{\sum_{\bf G,G^{'}{\neq0}}} G_{\alpha} \, v_{c}({\bf G})\,
v_{c}({\bf G^{'}}) \, G^{'}_{\beta} \\ 
{\times} \,
\int\,\psi^*_{\sigma{\bf k}}\,e^{i{\bf G}{\bf r}}\,
\psi_{\sigma^{'}{\bf k}}\,d\tau\,
\int\,\psi^*_{\sigma^{'}{\bf k}}\,e^{-i{\bf G^{'}}{\bf r^{'}}}\,
\psi_{\sigma{\bf k}}\,d\tau^{'}\, .
\end{multline}
%

The remarkable property of Eq.~(\ref{softmode-fr}) is that
it can be represented as a product of two independent factors,
${({\lambda}^{j})^{2}}\,\times\,\bar{{\Pi}}_{0}(0)$, one of which,
$({\lambda}^{j})^{2}=
{{\vert\,{g}^{j}\,\vert}^{2}\,[({\epsilon}_{\infty}+2)/3\,]^{-1}}\,$,
characterizes an effective interaction of electrons with polar zone-centre TO vibrations.
The other, $\bar{{\Pi}}_{0}(0)$, is the so-called non-interacting susceptibility.
Following the vibronic theory prescription (e.g., \cite{Kristoffel1}),
this allows us to constitute a reduced one-parameter version
of the el-TO-ph Hamiltonian~(\ref{hamiltonian_el-ph}) in the following model form:
\begin{equation}\label{hamiltonian_el-ph-fr1}
H_{el-ph}^{(red)} =
{N^{-1/2}} {\sum_{\sigma \neq \sigma^{'} }}\,{\sum_{\bf k}}\,
{{\lambda}^{j}_{{\sigma}{\sigma^{'}}}({\bf k})} \,
a^+_{\sigma{\bf k}}a_{\sigma^{'}{\bf k}} \, u_{0j} \,.
\end{equation}

First-principles justification of Eq.~(\ref{hamiltonian_el-ph-fr1})
is an important result of the present work.
We refer to this equation as a vibronic-type model since it involves 
the interband coupling between electrons and the zone-centre TO phonons.
According to~\cite{Kristoffel1,Kristoffel2}, such coupling forms the basis
of the vibronic theory.
Due to the successful mapping of the original microscopic model~(\ref{hamiltonian_el-ph})
onto the effective model given by~(\ref{hamiltonian_el-ph-fr1}),
it became clear
why, by introducing the constants ${\lambda}^{j}$ as model parameters, 
it is, in principle, possible within the framework of the vibronic theory
to study the lattice and electronic properties of polar materials.
Besides, Eq.~(\ref{softmode-fr}) shows the new meaning of the el-TO-ph coupling
constant, the square of which is inversely related to
the local field factor $({\epsilon}_{\infty}+2)/3\,$. 
Since local fields and their effects become more prominent in compounds with
a mixed ionic-covalent character of chemical bonds~(e.g.,~\cite{Dolgov,Resta,Katarzyna}),
this would certainly imply the important improvement in the applicability
of the standard vibronic Hamiltonian.
Note that a similar renormalization occurs
in the theory of surface polarization modes~\cite{Licari}.

From a physical point of view, Eq.~(\ref{hamiltonian_el-ph-fr1})
represents the Fr\"{o}hlich-type long-range electron-lattice dynamic hybridization
of the electronic bands of opposite parities, which takes into account the relevant
$s$-, $p$- and $d$-channels (e.g.,~\cite{Kristoffel1,Kristoffel2,Bersuker1,Bersuker2}
and references therein).
For instance, as applied to a family of high polar crystals such as 
the ferroelectric $\rm ABO_{3}$ perovskite oxides, hybridization~(\ref{hamiltonian_el-ph-fr1})
involves significant mixing between the filled $\rm O$ $\rm 2p$ and the empty ${\rm d}^{0}$
(${\rm Ti}^{4+}$, ${\rm Nb}^{5+}$, ${\rm Zr}^{4+}$, ${\rm Ta}^{5+}$, ${\rm Mo}^{6+}$,
${\rm W}^{6+}$, etc.) electronic states caused by the IR-active TO ${\rm F}_{1u}$ soft
vibrations~\cite{Kristoffel2,Konsin3,Konsin1}.

For the last several decades, the model of type (\ref{hamiltonian_el-ph-fr1})
was a useful prototype for intensive studies of various properties of both typical insulating
perovskites~\cite{Kristoffel2,Girshberg1,Konsin1,Ohnishi,Hidaka,Bussmann-Holder2}
and the $\rm A^{IV}B^{VI}$ narrow-gap semiconductors and their
alloys~\cite{Kawamura,Murase,Konsin2,Sakai,Maksimenko}. It was recently demonstrated
to be applicable to give a theoretical description of the ferroelectricity found in 
$\rm BiFeO_{3}$-type multiferroics~\cite{Konsin3,Konsin4}.
%
\subsection{Evaluation of electron-TO-phonon coupling at $q=0$}
%
In order to evaluate the el-TO-ph coupling in the long wavelength limit,
we first examine the Fr\"{o}hlich-type expression
for the amplitude of the interband el-TO-ph interaction
$g_{{\sigma}{\sigma}^{'}}({\bf k},{\bf q}j)$.
This can be accomplished
in a similar way as in the case of the scattering of an electron
due to LO phonons~\cite{Vinecki}.
Let us fix~$j$. Denoting the amplitude by
$
\vert \, g_{{\sigma}{\sigma}^{'}}({\bf k}) \vert =\,
(\lim_{{\bf q}\rightarrow0}\,{\vert \, g_{{\sigma}{\sigma}^{'}}({\bf k},{\bf q}j) \vert}^2 )^{1/2}
$
and using the evaluation of the dipole oscillator strengths of the TO modes
$S_{\alpha \beta}(j)$, we can obtain:
\begin{equation}\label{el-TO phonon1}
\vert \, g_{{\sigma}{\sigma}^{'}}({\bf k}) \vert
\,\cong\,
{\Delta{\Omega}}_{LT}\,
\sqrt{\, \frac{ v \, {{\epsilon}_{\infty}} }
{4\pi} \, } \,
\left| <{\sigma}{}{\bf k} |
{\bf F}_{0}({\bf r})
| {\sigma}^{'}{\bf k}> \right|
\end{equation}
where, according to~\cite{Pickett}, we introduced for the given polarization
a difference between the longitudinal (${\Omega}_{LO}$) and
the transverse (${\Omega}_{TO}$) frequencies (the LO-TO splitting)
provided in terms of
${\Delta{\Omega}}_{LT} = ( {\Omega}^{2}_{LO} - {\Omega}^{2}_{TO} )^{1/2} $
(i.e., the difference in long-range fields given by 
longitudinal and transverse modes as ${\bf q}\rightarrow0$).

From Eq.~(\ref{el-TO phonon1}), one can draw the following important inferences:
First of all, note that dynamic hybridization described by~(\ref{hamiltonian_el-ph})
leads to an asymmetric charge distribution,
which corresponds to the internal electric field ${\bf F}_{0}({\bf r})$. This field,
in turn, provides, due to Eq.~(\ref{transversefield1}), the long-range character
of the el-TO-ph coupling at the $\Gamma$ point.
Secondly, the magnitude of the splitting ${\Delta{\Omega}}_{LT}$
serves as an enhancement factor of the matrix elements of ${\bf F}_{0}({\bf r})$.
Therefore, in the long-wavelength limit, the el-TO-ph interaction in polar crystals
is distinguished from that of most other dielectrics by the following features:
(i) it is long-range, (ii) it is controlled by the internal electric field
and (iii) it is essentially sensitive to values of the LO-TO difference
(i.e., the polar strength of long-wavelength optical vibrational modes).
Eq.~(\ref{el-TO phonon1}) can serve as indicative of the strength
of the el-TO-ph coupling in polar materials. In particular,
large values of $\vert \, g_{{\sigma}{\sigma}^{'}}({\bf k}) \vert$
occur when ${\Delta{\Omega}}_{LT}$ is large:
for example, the experimental difference 
between the corresponding LO and TO modes in the ferroelectric
$\rm BaTiO_3$ is about 530 $\rm cm^{-1}$~\cite{Kuo}
to be compared with the values of 100, 69, and 48 $\rm cm^{-1}$
obtained for the differences in the compounds
$\rm NaCl$, $\rm KCl$, and $\rm RbCl$, respectively~\cite{Raunio}.

In a polar lattice, the LO-TO splitting depends generally on Born{\textquoteright}s
transverse dynamical effective charge $Z^{*}$ of the lattice ions and
the screening of the Coulomb interaction, which depends on the electronic
dielectric permittivity ${\epsilon}_{\infty}$~\cite{Zhong}:
${\Delta{\Omega}}_{LT}^{2} \propto \left|Z^{*}\right|^{2}/{\epsilon}_{\infty}\,$.
Thus, one can conclude that the presence of anomalously large Born{\textquoteright}s
effective charges is the key signature that the interband el-TO-ph coupling is
essentially strong in a ferroelectric material. 
This result corresponds to the main assumption of the vibronic
theory~\cite{Kristoffel1,Kristoffel2,Bersuker1,Bersuker2} regarding the existence of
sufficiently strong interband el-TO-ph coupling in displacive ferroelectrics.
The strength of the interband el-TO-ph interaction
can therefore serve as a direct indicator of the extent to which a crystal lattice is close
to a possible ferroelectric instability.
Further details concerned with the quantitative characteristics of
the electron-TO-phonon coupling are given in our separate paper~\cite{Pishtshev1}.
%
\subsection{Electron-TO-phonon coupling via macroscopic parameters}
%
As shown in the present work, one can associate the interaction between electrons
and the polar long-wavelength TO phonons with the long-range dipole-dipole interaction.
This link, which was not anticipated by previous theoretical considerations,
helps us to express the strength of the el-TO-ph interaction 
via macroscopic parameters of a polar crystal.

Let us introduce the bare constant $g_{0j}$ which represents
the effective ${\bf k}$-independent interband
el-TO-ph interaction defined by a proper average of the squared el-TO-ph matrix
elements over the given electronic states:
\begin{equation}\label{bare_el-TO-ph}
g_{0j}^{2}=
{{\sum_{\sigma^{'}{>}\,\sigma}} \,
\frac{1}{2N} 
{\sum_{\bf k}}} \,
{\vert \, g^{j}_{{\sigma}{\sigma^{'}}}({\bf k}) \vert}^2\,
\frac{{E}_g}{ \,\left|\,{E_{\sigma^{'}}({\bf k})-E_{\sigma}({\bf k})} \right|\, } 
\end{equation}
where ${E}_g$ is a bond-gap energy. Accounting for local-fields effects-induced
partial screening of the bare el-TO-ph interaction is described by the renormalization:
\begin{equation}\label{renorm}
{g}_{0j}^{2}\,\longrightarrow\,{\bar{g}}_{j}^{2}\,=\,
{3{g}_{0j}^{2}}/{({\epsilon}_{\infty}+2)}\,.
\end{equation}
As follows from Eqs.~(\ref{softmode-eq3}) and (\ref{softmode-eq3-B1}),
the contribution of the linear el-TO-ph interaction
to the square of the zone-centre TO vibrational mode frequency $\Omega_{0j}^{2}$
can be represented as the product of two macroscopic quantities:
the quantity $ B_{\alpha \beta}(0)$ which accounts for the corresponding
electronic contribution,
and the dipole oscillator strengths $S_{\alpha \beta}(j)$ associated with
the given TO vibration.
The bare constant of the el-TO-ph interaction $g_{0j}$
can thus be determined from matching Eq.~(\ref{softmode-fr}) 
and Eq.~(\ref{softmode-eq3}) together.
As a result, we obtain the following relationship
($S(j)={\sum}_{\alpha}\,S_{\alpha \alpha}(j)\,$):
\begin{equation}\label{ephto0}
g_{0j}^{2} \,=\, \frac{1}{12} {{E}_g} \, M_{j} \, S(j) \, .  
\end{equation}

Eq.~(\ref{ephto0}) is an important result of the present work.
It relates, for each zone-center TO vibration 
of the branch $j$ and of the reduced mass $M_j$,
the bare constant of the el-TO-ph coupling at the $\Gamma$ point,
$g_{0j}$, with the macroscopic material constants, the values of which 
can be obtained from experiment.
Employing the relevant information from the IR spectra
concerning the IR-active optical phonons and the dielectric function behavior
in the far-IR spectral range, and using Eqs.~(\ref{ephto0}) and~(\ref{renorm})
in combination with
the experimental data for ${\epsilon}_{\infty}$ and ${E}_g$, we can directly evaluate
the el-TO-ph coupling constants for polar materials of interest.

A more detailed demonstration of the practical usefulness of Eq.~(\ref{ephto0}),
together with numerical results for a wide number of selected polar insulators
and semiconductors, can be found in our separate publication~\cite{Pishtshev1}.
In particular, calculations of the interband el-TO-ph interaction constants
and the further comparative analysis showed
that the large interband el-TO-ph interaction is a special microscopic feature
of the ferroelectric materials. In contrast, in non-ferroelectrics, as it was
demonstrated, the strength
of the el-TO-ph interaction is not necessarily high enough due to their lower polar nature.
%
\section{Conclusions}
In this paper, focusing mostly on the fundamental contribution of electron subsystem
to the dynamics of polar long-wavelength TO vibrations and using a first-principles methodology, 
we provided a systematic description of the el-TO-ph interaction in a polar insulator.
The study is based on the model of a polar crystal with classical potentials,
which takes into account the electronic polarizability effects.
By analyzing the electronic contribution to the TO vibrational mode in terms of the el-TO-ph
coupling, we established a bridge which allowed us
a) to link the model under consideration to the microscopic lattice dynamics and
b) in the long wavelength limit, to relate the interaction of electrons
with polar TO vibrations to the long-range dipole-dipole interaction.
Our results highlight the importance of the el-TO-ph interaction
for the genesis of the long-wavelength TO vibrations in ferroelectrics,
thereby giving fundamental support at the microscopic level
for the applicability of the vibronic theory.
Within a first-principles methodology, we found and explained the significant increase
of the constants of the el-TO-ph coupling in ferroelectric materials,
showed how the el-TO-ph interaction constants might be dependent on macroscopic material
parameters, and obtained analytical equations allowing us to estimate el-TO-ph interaction
strengths in a wide range of polar dielectrics.
In particular, it was proved that the zone-center TO vibrational mode effective charge $Z^{*}$
can be considered to be the key macroscopic parameter of the el-TO-ph coupling strength.
In materials where the el-TO-ph coupling is operative,
it can be verified by spectroscopy measurements of the IR-active TO mode.
%
\section*{Acknowledgements}
The author would like to thank N.~Kristoffel for helpful discussions,
A.~Sherman for attention to this work.
The work was supported by the Estonian Science Foundation grants No. 6918
and No. 7296.
%
%

%
\end{document}